\documentclass[letter]{aa}     

\usepackage{graphicx}
\usepackage{txfonts}
\usepackage{lipsum}
\usepackage{subcaption}         
\usepackage{lscape}             
\usepackage{placeins}           
\usepackage{xcolor}
\usepackage{hyperref}
\newcommand{\gC}{$\gamma$\,Cas\xspace}
\newcommand{\Cy}{59\,Cyg\xspace}
\begin{document}


\title{HD 120678: a classical Be star with a tilting heartbeat disk}

%
%
%

   \author{D. Baade\inst{1}\corrauth{dbaade@eso.org}        
   \and
   J. Labadie-Bartz\inst{2,3}\email{jbartz@udel.edu}   
   \and 
   Th. Rivinius\inst{4}\email{triviniu@eso.org}
}

   \institute{European Organisation for Astronomical Research in the Southern Hemisphere (ESO), Karl-Schwarzschild-Str. 2, \\ 85748 Garching b. M\"unchen, Germany
   \and
   DTU Space, Technical University of Denmark, Elektrovej 327, Kgs. Lyngby 2800, Denmark
   \and
LIRA, Paris Observatory, CNRS, PSL University, Sorbonne University, Universit{\' e} Paris Cit{\' e}, CY Cergy Paris University, 5 place Jules Janssen, 92195 Meudon, France
   \and
   European Organisation for Astronomical Research in the Southern Hemisphere (ESO), Casilla 19001, Santiago 19, Chile
}

\date{Received ??}

\abstract{
Classical Oe/Be/Ae stars possess Keplerian circum-equatorial decretion disks which produce not just the eponymous emission lines but also the bewildering diversity of photometric variations on timescales above a few days.  The variability is due to varying amounts and geometric distributions of the circumstellar gas.  The ASAS-3 $V$-band light curve of \object{HD\,120678} between March 2005 and November 2008 resembled an inverse P Cygni profile, which may have been a one-off event.   The peak-to-valley amplitude reached $\sim$0.9\,mag.  Before and after this photometric event, the Balmer emission lines and their variability matched those of classical Be stars with an intermediate to low inclination angle.  At the sharp minimum of the light curve, \citet{2012A&A...546A..92G} observed shell-type profiles implying an edge-on (high-inclination) view.  For the historical activity phases of \object{\gC} and \object{59\,Cyg}, such a Be => Be-shell => Be sequence has been successfully interpreted as rotations of the disk plane.  \object{HD\,120678} and a few other Be stars add a not previously recognized phenomenon to the plethora of Be-star variabilities, namely heartbeat light curves.  Analogously to the distortion of heartbeat stars, the variability is due to heartbeat disks gravitationally tilted by an additional star as modeled by \citet{2022ApJ...928..145M} for Pleione.  While the typical amplitudes of heartbeat stars are only a few 0.01\,mag, heartbeat disks can explain the ten times larger amplitudes observed in Be stars.  
}


\maketitle
\nolinenumbers

\section{Introduction}
\label{sect:intro}
Latest with the interferometric resolution of disks \citep{1997ApJ...479..477Q} around classical Be stars \citep{2013A&ARv..21...69R, 2026enap....2..430R}, Struve's (\citeyear{1931ApJ....73...94S}) epochal paradigm of a gaseous, flat, equatorial, and rotationally supported circumstellar structure has become established.  Models very satisfactorily reproduce the observed landscapes of spectral energy distributions \citep{2017A&A...601A..74K} and emission-line profiles \citep{2025Ap&SS.370..120R}.  Time-dependent mass loss from the central B star is a challenge but does not pose fundamental problems \citep{2021ApJ...909..149G}.  Nevertheless, models occasionally fail catastrophically.  In some cases, this is due to the built-in assumption of an invariant disk plane, which is the traditionally unchallenged heritage of Struve's concept.  

The historically most prominent problem cases are \gC and \Cy, which were attributed to tilting disks only decades later \citep{1998A&A...330..243H}.  \citet{2023A&A...678A..47B} added to this ansatz that disks can bend and rotate from the equator over the stellar poles, and \citet{2022ApJ...928..145M} found that the disk in Pleione regularly tilts and tears.  Isolated Be stars are incapable of producing large-scale disk motions.  The hot low-mass companions to many early-type Be stars are the remains of a former mass and angular-momentum donor \citep{2025A&A...698A..38B}.  Therefore, they are probably on orbits that are not sufficiently misaligned to exert enough torque so that putative tertiaries may have to be considered \citep{BaadeLiege}.  

The literature and public databases contain more examples of Be stars that have exhibited symptoms of a migrating disk plane.  This Letter identifies HD\,120678 as kind of a "Rosetta star" as it combines the spectroscopic indications of variable disk planes known from other Be stars \citep[][]{2023A&A...678A..47B, BaadeLiege} with a "heartbeat" light curve (LC), a newly recognized variability pattern of Be stars.  After an introduction to heartbeat stars in Sect.\,\ref{sect:heartbeats},  Sect.\,\ref{sect:literature} summarizes the literature on HD\,120678.   The evidence of heartbeat disks in this and some other classical Be stars is screened in Sect.\,\ref{sect:other}, which also includes a comparison to \object{MACHO\,80.7443.1718}, a B0Iae supergiant in a 33-d binary with a disk and a strong heartbeat LC.  A summary and the conclusions follow in Sect.\,\ref{sect:conclusions}.

\section{Heartbeat stars}
\label{sect:heartbeats}

Because of a perceived similarity of some stellar LCs to human electrocardiograms, \citet{2012ApJ...753...86T} coined the term "heartbeat stars" for ellipsoidal variables on a strongly elliptical orbit.  Today, more than a thousand heartbeat systems are known \citep{2022ApJS..259...16W}.  Their low $g$ makes the primaries deformable: 90\% are giants and 10\% in the Hertzsprung gap.  Typical orbital periods range from a week to a year.  

The main component of the variability is due to gravitational distortion around periastron.  In systems with hot components, heating by mutual irradiation also becomes important.  The sensitivity of both effects to orbital inclination, eccentricity, and argument of periastron can be used to estimate these parameters \citep{2020ApJ...903..122C}.  The combined amplitudes are typically a few 0.01\,mag, and the LCs are diverse \citep{2022ApJ...928..135W, 2022ApJS..259...16W}.  Depending on the orientation of the orbits, they can be symmetric with one or two peaks.  Others look like spectroscopic P~Cygni or inverse P Cygni profiles.  Furthermore, $\sim$15\% of the heartbeat stars exhibit tidally excited oscillations \citep[TEOs;][]{2020ApJ...903..122C} with frequencies of an integer multiple of the orbital frequency and amplitudes mostly at the sub-mmag level.

\section{HD 120678 in the literature}
\label{sect:literature}

\citet{1977ApJ...213..438C} published a $v$\,sin\,$i$ of 350\,km\,s$^{-1}$, which would suggest a moderate to high inclination.  More recently, \citet{2024A&A...692A.192B} reported a drastically lower value of 120\,km\,s$^{-1}$ 
from \ion{O}{III}\,$\lambda$5592.  However, inspection of the FEROS spectra used by \citet{2012A&A...546A..92G} and probably by \citeauthor{2024A&A...692A.192B}, too, shows that variable emission components can make this line unsuitable for estimating $v$\,sin\,$i$.  The base width of all non-hydrogen lines is 400-500\,km\,s$^{-1}$, implying v\,sin\,$i$\,$\leq$\,250\,km\,s$^{-1}$ and an intermediate to low inclination of this O9e star.  

The spherical-symmetry-breaking disks of Be stars imprint a linear polarization on the light \citep{1975ApJ...196..777P, 2010ApJ...709.1306W}.  It vanishes for circularly symmetric disks viewed face-on.  \citet{2009A&A...505..743V} found \object{HD\,120678} unpolarized in 2003 and saw "no notable difference" to earlier measurements.  he H$\alpha$ flux profile had a single sharp peak.  Its spectral resolution was left undocumented but \citeauthor{2009A&A...505..743V} presented double emission peaks for other Be stars.  A FEROS H$\alpha$ spectrum from 2012 with $R$\,=\,46,000 as a lower peak but identical shape.  Therefore, the disk inclination angle was low to intermediate in 2003 (and before) and again in 2012, crudely bracketing the shell-like event in 2008 (Sect.\,\ref{sect:HDspectral}).  Comparison to model emission-line profiles \citep{2020ApJ...894...18S} also suggests an intermediate inclination, which presumably is the equilibrium position of the disk.

\subsection{Spectral variations}
\label{sect:HDspectral}

From radial-velocity (RV) measurements of two \ion{He}{ii} absorption lines in 32 optical spectra spanning 5 years, \citet{2012A&A...546A..92G} derived a tentative period of 280\,d but did not publish their RV curve.  The amplitude was of order 20\,km\,s$^{-1}$.  The H$\beta$ to H$\delta$ emission profiles at the beginning and the end of the sequence were relatively narrow with a weak central reversal matching a low to intermediate inclination \citep[as expected for an equatorial disk around a star with this inclination;][]{2010ApJS..187..228S}. 

In a single night in June 2008, \citeauthor{2012A&A...546A..92G} found a shell-like event marked by the transient presence of absorption features in Balmer, \ion{He}{i}, and singly- and doubly-ionized metal lines.  With $\sim$3\,\AA\ ($\approx$\,200\,km\,s$^{-1}$ at resolving power $R$\,$\approx$\,15,000), the width of the shell components does not fit the conventional properties of shell lines in Be stars \citep[cf.\,][]{2006A&A...459..137R}.  The lines concerned were of lower excitation than the photospheric lines, which remained $\sim$unchanged and resembled an early-B spectral type.  Balmer and helium line emission weakened strongly or disappeared, and the peak separation in H$\beta$ and \ion{He}{I} $\lambda$5875 more than tripled.  In the two preceding years, the violet-to-red (V/R) flux ratio in the emission peaks of H$\beta$ and H$\gamma$ varied.  During the shell phase, it was $\approx$1, characteristic of an axisymmetric disk.

The spectra closest in time to the event are from one month before and one year after it.  Because of their lower $R$\,$ \approx$\,2,500, there is not much structure in the Balmer emission profiles.  Since the H$\beta$ emission was well above the continuum at all times, a direct transition from a Be to a Be-shell-like phase occurred within one month.  The last spectrum before the event showed hints of the appearance of absorption lines, and the width of the H$\beta$ emission was already close to it maximum.

\subsection{Photometry}
\label{sect:HDphot}

The well-sampled ASAS $V$-band LC \citep{2012A&A...546A..92G} resembles an inverse P Cygni profile as in many heartbeat binaries.  Starting in 2005, the brightness increased in an accelerating fashion by 0.4\,mag from a plateau to a narrow peak in 2008.  The descent by 0.85\,mag to an even narrower minimum took just 160\,d, after which the star returned within 120\,d to a level slightly above that in 2005.  The time of the deep minimum coincided with that of the observation of the shell event to within $\sim$1~day.  In the ASAS \footnote{\url{https://www.astrouw.edu.pl/asas/?page=aasc}} and AAVSO\footnote{\url{https://apps.aavso.org/v2/tools/light-curves/}} LCs, there are $\sim$3 further perturbations but without obvious heartbeat pattern and any regularity in their spacing in time.   In a single photoelectric observation between 1971 and 1973, \citet{1975A&AS...19..243H} measured $V$\,=\,7.62\,mag, $\sim$0.2\,mag brighter than the 2001-2009 plateau and, therefore, possibly suggestive of an earlier disk tilt.

\citet{2004AcA....54..153P} published an automatically determined photometric period of $\sim$272\,d, based on ASAS data from 2001 to 2004.  In the more complete ASAS LC \citep{2012A&A...546A..92G}, fluctuations by up to 0.2\,mag can be seen with an average spacing of $\sim$330\,d.  Since this is 3-4 times shorter than the base width of the heartbeat, the latter was presumably not due to a putative companion's one-year orbit alone.

\subsection{Discussion}
\label{sect:discuss}
The low polarization, which is in agreement with the contemporaneous narrow H$\alpha$ emission in 2003, and the temporary tripling of the emission-peak separations with embedded narrow absorptions in 2008 trace variations in the inclination of HD\,120678 spanning a range of $\sim$45$^{\circ}$.  The conflicting stellar $v$\,sin\,$i$ measurements mean that the equilibrium inclination of the disk relative to the equator is spectroscopically poorly constrained. 

The shape of the LC is unusual but it is only beginning to be recognized as a distinct variability pattern of Be stars (see below).  The immediate importance of HD\,120678 lies in it combining the spectroscopic evidence of a migrating disk with a heartbeat LC to the concept of a heartbeat \underline{disk}.  

It was fortunate that \citet{2012A&A...546A..92G} observed the shell-like features within just one day of the photometric minimum.  This makes this moment the pivotal point for the understanding of the variability: both photometry and spectroscopy imply that the outer disk had an edge-on inclination of $\sim$90$^{\circ}$.  These observations are not a statistical fluke: one month before the shell event, the brightness was already close to the minimum, and the width of the H$\delta$ emission was near its maximum.  

Before and after the event, the brightness was the same and $\sim$constant.  The same holds for the width of the emission lines, suggesting that the disk has an equilibrium position, which in Be stars is centered on the equator \citep{1931ApJ....73...94S, 1997ApJ...479..477Q, 2006A&A...459..137R}.  The variability of the disk around HD\,120678 can, therefore, be described as follows: 
\vspace{-2mm}

\begin{enumerate}
\item 
The unperturbed disk was in an equatorial location. 
\item 
The disk plane gradually turned such that the projected surface increased and the observer received more light.  
\item 
The disk plane reached its position closest to face-on, and the brightness was at maximum. 
\item
The disk turned back, and crossing the equator, moved on to the edge-on view when the shell lines appeared and the system was faintest because the visible disk area was smallest. Any "overshooting" is observationally unconstrained.
\item 
The disk returned to the equatorial equilibrium position.  
\end{enumerate}
\vspace{-2mm}

The simultaneity of photometric minimum and shell episode is the consequence of the edge-on view of the disk.  There is no reason, though, to believe that all Be stars with a heartbeat disk develop shell lines at the photometric minimum.  However, if they do, the brightness should be lowest because the visible disk area is smallest (and the extinction of stellar light largest).

\section{Other stars with symptoms of heartbeat disks}
\label{sect:other}

\subsection{Pleione}
\label{sect:Pleione}
Pleione exhibits one of the most complex, but regular variability patterns known of Be stars.  In the elaborate model by \citet{2022ApJ...928..145M}, the varying tilt of the disk is the main contributor to the photometric cycles and also produces intermittent shell phases.  At intervals of 32 yr, the outer disk is gravitationally torn off by the companion on an eccentric misaligned 218-d orbit and afterwards precesses freely about the star but dissipates because it is cut off from the mass supply from the star.  The inner disk returns to its equatorial equilibrium position where it is fed.  

\subsection{88 Her}
\label{sect:88Her}
The visual LC of the SB1 Be star \object{88\,Her} \citep[with P\,=\,86.7\,d,][]{1972ApL....11..119H} faded four times by $\sim$0.25\,mag over $\sim$500\,d and returned to its presumable equilibrium brightness much more slowly 
\citep{2026arXiv260706104B} 
in a fashion very similar to Pleione.  During the steep decline phases of the LC, the shell spectrum got much stronger.  \citet{Haubois2012} attributed the events to outbursts.  The similarity of the $V$-band profiles of the four well separated events identifies them as template of such events.  The photometric observations distinguishing Pleione and 88\,Her are their \textit{U-B} color curves, illustrating the importance of color information for the understanding of the variability of Be disks.  More critical for the comparison of the two stars is the only transient presence of shell absorptions in Pleione and their (quasi-)permanence in 88\,Her, showing that the disk of 88\,Her does not move but more probably varies in thickness.

\subsection{AzV 493}
\label{sect:AzV 493}
The observed LC of this O3-O5e star has a peak-to-valley amplitude of $\sim$1.2\,mag and may consist of two 7.3-yr cycles \citep{2023ApJ...947...27O, 2026PASP..138g4202O}.  The first cycle was clearly of the heartbeat type, the beginning of the second one was severely perturbed by additional oscillatory variations, and the third one started like the first.  If the Be+sdO/B or Be+WD paradigm with past mass exchange is applicable to AzV\,493 and the 7.3/14.6-yr cycle is due to a companion, AzV\,493 would be a triple system.

\subsection{$\psi$ Per}
\label{psiPer}
The multiplicity of this late-type Be star is unclear \citep{2023Galax..11...83M}.  In agreement with the persistent shell absorption, \citet{2011A&A...529A..87D} derived $i$\,=\,75\,$\pm$\,8$^{\circ}$ from their interferometry.  In the AAVSO LC with CCD $B$- and $V$-band magnitudes, a heartbeat event with inverse P Cyg-like light profile and a peak-to-valley amplitude of $\sim$0.5\,mag in $V$ is indicated from mid-2018 to mid-2022.  Visual inspection of numerous echelle spectra in the BeSS database \citep{BeSS} observed during, before, and after this time interval did not identify any appreciable changes in Balmer and metal emission lines or shell absorptions.  This stability is typical of disks around late-type Be stars, and the heartbeat may indicate changes in vertical scale height rather than inclination.

\subsection{MACHO 80.7443.1718}
\label{sect:MACHO}
This SB1 star in the LMC with a period of 32.8\,d and an eccentricity of 0.51 \citep{2021MNRAS.506.4083J} is not a classical Be star, but it provides for an interesting comparison.  On account of the amplitude of the stable P Cyg-shaped LC of up to 50 times that of many heartbeat stars (0.5 vs.\ a few 0.01\,mag), \citeauthor{2021MNRAS.506.4083J} dubbed \object{MACHO\,80.7443.1718} the "loudest heartbeat star".  Noting the emission lines, they called it a B[e] as well as a B0\,Iae star.  However, without evidence of dust or forbidden emission lines \citep[see also][]{2024A&A...686A.199K}, the basis for the classification as a B[e] star \citep{2019Galax...7...83K} is missing.  

\citet{2021MNRAS.506.4083J} reported the detection of three TEO frequencies, and \citet{2022A&A...659A..47K} added two more, establishing strong tidal effects.  
Nevertheless, an overlapping team of authors, \citet{2024A&A...686A.199K}, considered tidal distortion of only secondary importance as an explanation of the main photometric variability, using the achromaticity of the \textit{UBV} heartbeat amplitudes as the main argument.

A fair amount of discussion adopting the heartbeat-binary interpretation \citep[e.g.,][]{2024A&A...685A.145K, 2022A&A...659A..47K} followed the discovery report.  \citet{2023NatAs...7.1218M, 2026ApJ..1006..158M} attributed the LC to spectacular breaking waves on the surface of the primary.  By reference to \object{$\delta$\,Sco}, \citet{2021MNRAS.506.4083J} suggested outbursts occurring after each periastron passage.  However, today's consensus is that the change in effective gravity of $\sim$1\% produced by the passing companion to $\delta$\,Sco is too small to cause significant mass loss \citep[][see also \citeauthor{2024ApJ...968...30R} \citeyear{2024ApJ...968...30R}]{2013ApJ...766..119M}.  Heartbeats are not known of $\delta$\,Sco.  

It is instructive to compare MACHO\,80.7443.1718 to the so far most massive heartbeat star \citep[\object{$\iota$\,Ori},][]{2017MNRAS.467.2494P}.  The published spectral types are B0\,Ia or B0[e] for the former and O9\,III for the latter.  The orbital periods of these two SB1 stars are 32.8\,d vs.\ 29.1\,d, the orbital inclination angles 43.9\,$\pm$\,0.2$^{\circ}$ 
vs.\ 62.86$^{+0.17}_{-0.14}$$^{\circ}$, and the eccentricities are 0.51 and 0.76, respectively.  Yet, the amplitude of MACHO\,80.7443.1718 is 0.5\,mag while that of $\iota$\,Ori is $\sim$0.01\,mag \citep[all numbers are from][and \citeauthor{2017MNRAS.467.2494P} \citeyear{2017MNRAS.467.2494P}, respectively]{2021MNRAS.506.4083J}.  This suggests that  the variabilities of the two stars do not belong to the same league.  

The key to the understanding of the nature of MACHO\,80.7443.1718 probably lies in the geometry of the line-emitting region.  It must be rotating rapidly because the separation of the two emission peaks varies between $\sim$300 and $\sim$450\,km\,s$^{-1}$ \citep{2021MNRAS.506.4083J, 2024A&A...686A.199K}.  It may surround the primary although co-phased RV curves of the primary and the presumable disk do not seem to have been published.  \citet{2024A&A...686A.199K} did not achieve a satisfactory fit of the phase dependence of the H$\alpha$ and H$\beta$ profiles.  However, they assumed that the disk is coplanar with the primary's equator.  By contrast to this star, a disk is not known in $\iota$\,Ori, where the H$\alpha$ emission forms in the wind with small orbital variations \citep{1996ApJ...469..884G}. 

Tenuous disks are more readily distorted than stars, and model calculations demonstrate a large aspect-angle sensitivity of their brightness \citep{Haubois2012, 2023MNRAS.526..782S}.  The high similarity to the photometric variability of HD\,120678 suggests to seek the explanation of the heartbeats of MACHO\,80.7443.1718 in its circumstellar disk although, owing to the absence of shell as well as single-peak profiles, the tell-tale signatures of disk migration are missing.  Only simultaneous modeling of the emission-line profiles and the light and color curves can reveal whether or not there is a cyclically tilting, rotating disk about the primary in \object{MACHO\,80.7443.1718}.  

High-luminosity OB stars are not generally known to develop circumstellar disks \citep{2009A&A...505..743V}, although there are exceptions like the B1Ib star $\gamma$ Ara \citep{1983A&A...124..211B}.  However, in MACHO\,80.7443.1718, a disk might be due to the fast rotation of the primary ($247\pm48$\,km\,s$^{-1}$) which is near the maximum observed in blue supergiants \citep[cf.\,][]{2014A&A...562A..37M}, and further assisted by the secondary.  Model calculations suggest that reverse mass transfer is more likely for very short periods \citep{2025A&A...699A.178Q} and so eliminate this process as an alternative explanation.  Compared to classical early-type Be stars, the H$\alpha$ emission is relatively weak and unusually stable.  This could be the growth-restricting effect of the tight elliptical orbit of the companion, which may every month shave off the outer disk.

Finally, \citet{2022A&A...659A..47K} analyzed the stability of the period and found a decrease by 11\,s\,yr$^{-1}$.  As the most likely explanation they considered a third star with a mass of $\sim$18\,M$_{\odot}$ on a $\gtrsim$30-yr orbit.

\section{Summary and conclusions}
\label{sect:conclusions}

The variable Balmer line profiles of HD\,120678 demonstrate that the disk assumed different inclination angles.  The single shell phase was coincident in time with the only confirmed photometric heartbeat.  Therefore, the heartbeat, too, was due to a transient tilting/bending of the disk.  The spectroscopic shell phases and photometric minima of Pleione are also roughly synchronized \citep{2022ApJ...928..145M}.
In $\psi$\,Per, no spectroscopic counterpart to the apparent photometric heartbeat exists. The non-Be star MACHO\,80.7443.1718 may illustrate that high-amplitude heartbeats are a disk phenomenon not restricted to Be stars.  

For all heartbeat Be stars above, an understanding of the large time difference between the events is lacking, which in HD\,120678 (and $\psi$\,Per) is unconstrained.  \citet{2024MNRAS.527.7515S} predict some regularity in the timing of disk-tearing events in misaligned two-star systems.  But a genuine 34-yr periodicity in Pleione would require a clock that perhaps only a third star can provide.  
In AzV\,493, too, there is explanatory scope for a third star \citep[as proposed for MACHO\,80.7443.1718 by][]{2022A&A...659A..47K}.   That is, tertiary stars with misaligned and/or eccentric orbits could be very effective at triggering the out-of-equilibrium motions of the disk planes.   

Understanding heartbeat disks does not promise to add to the understanding of the genesis of Be disks, i.e., will not shed new light on the Be phenomenon.  However, since after each heartbeat the disks return to their equilibrium positions, there must be a restoring force.  This is most likely the viscosity so that more can be learned about it from studying also tilting heartbeat disks than just from in-plane disk structures.  Moreover, UV spectra of Be stars observed at times when their disks are substantially outside their equilibrium position may be able to probe the latitudinal dependence of their winds.  Yet, the most important point about heartbeat disks is the awareness of them.  Without considering the strong observational effects of migrating disks, drastic confusions are possible.  Conversely, in disks with persistent edge orientation, increased or intermittent shell absorption caused by mass-loss outbursts can be confused with tilting.


\begin{acknowledgements}
Please see Appendix \ref{acknowledgments}.
\end{acknowledgements}


\bibliographystyle{aa} 
\bibliography{dbaade} 

@ARTICLE{2024MNRAS.527.7515S,
       author = {{Suffak}, M.~W. and {Jones}, C.~E. and {Carciofi}, A.~C.},
        title = "{Disc tearing in a Be star: predicted 3D observations}",
      journal = {\mnras},
         year = 2024,
        month = jan,
       volume = {527},
       number = {3},
        pages = {7515-7522},
          doi = {10.1093/mnras/stad3659},
archivePrefix = {arXiv},
       eprint = {2311.14185},
 primaryClass = {astro-ph.SR},
       adsurl = {https://ui.adsabs.harvard.edu/abs/2024MNRAS.527.7515S}
}

@ARTICLE{2013A&ARv..21...69R,
   author = {{Rivinius}, T. and {Carciofi}, A.~C. and {Martayan}, C.},
    title = "{Classical Be stars. Rapidly rotating B stars with viscous Keplerian decretion disks}",
  journal = {\aapr},
  
archivePrefix = "arXiv",
   eprint = {1310.3962},
 primaryClass = "astro-ph.SR",
     year = 2013,
    month = oct,
   volume = 21,
      eid = {69},
    pages = {69},
      doi = {10.1007/s00159-013-0069-0},
   adsurl = {http://esoads.eso.org/abs/2013A%26ARv..21...69R}
}

@ARTICLE{2006A&A...459..137R,
   author = {{Rivinius}, T. and {{\v S}tefl}, S. and {Baade}, D.},
    title = "{Bright Be-shell stars}",
  journal = {\aap},
     year = 2006,
    month = nov,
   volume = 459,
    pages = {137-145},
      doi = {10.1051/0004-6361:20053008},
   adsurl = {http://esoads.eso.org/abs/2006A%26A...459..137R}
}

@ARTICLE{2010ApJ...709.1306W,
   author = {{Wisniewski}, J.~P. and {Draper}, Z.~H. and {Bjorkman}, K.~S. and 
	{Meade}, M.~R. and {Bjorkman}, J.~E. and {Kowalski}, A.~F.},
    title = "{Disk-Loss and Disk-Renewal Phases in Classical Be Stars. I. Analysis of Long-Term Spectropolarimetric Data}",
  journal = {\apj},
archivePrefix = "arXiv",
   eprint = {0912.1504},
 primaryClass = "astro-ph.SR",
     year = 2010,
    month = feb,
   volume = 709,
    pages = {1306-1320},
      doi = {10.1088/0004-637X/709/2/1306},
   adsurl = {http://adsabs.harvard.edu/abs/2010ApJ...709.1306W}
}

@ARTICLE{Haubois2012,
   author = {{Haubois}, X. and {Carciofi}, A.~C. and {Rivinius}, T. and {Okazaki}, A.~T. and 
	{Bjorkman}, J.~E.},
    title = "{Dynamical Evolution of Viscous Disks around Be Stars. I. Photometry}",
  journal = {\apj},
archivePrefix = "arXiv",
   eprint = {1207.2612},
 primaryClass = "astro-ph.SR",
     year = 2012,
    month = sep,
   volume = 756,
      eid = {156},
    pages = {156},
      doi = {10.1088/0004-637X/756/2/156},
   adsurl = {http://esoads.eso.org/abs/2012ApJ...756..156H}
}

@ARTICLE{2010ApJS..187..228S,
   author = {{Silaj}, J. and {Jones}, C.~E. and {Tycner}, C. and {Sigut}, T.~A.~A. and 
	{Smith}, A.~D.},
    title = "{A Systematic Study of H{$\alpha$} Profiles of Be Stars}",
  journal = {\apjs},
     year = 2010,
    month = mar,
   volume = 187,
    pages = {228-250},
      doi = {10.1088/0067-0049/187/1/228},
   adsurl = {http://cdsads.u-strasbg.fr/abs/2010ApJS..187..228S}
}

@ARTICLE{BeSS,
   author = {{Neiner}, C. and {de Batz}, B. and {Cochard}, F. and {Floquet}, M. and 
	{Mekkas}, A. and {Desnoux}, V.},
    title = "{The Be Star Spectra (BeSS) Database}",
  journal = {\aj},
     year = 2011,
    month = nov,
   volume = 142,
      eid = {149},
    pages = {149},
      doi = {10.1088/0004-6256/142/5/149},
   adsurl = {http://esoads.eso.org/abs/2011AJ....142..149N}
}

@ARTICLE{2017MNRAS.467.2494P,
   author = {{Pablo}, H. and {Richardson}, N.~D. and {Fuller}, J. and {Rowe}, J. and 
	{Moffat}, A.~F.~J. and {Kuschnig}, R. and {Popowicz}, A. and 
	{Handler}, G. and {Neiner}, C. and {Pigulski}, A. and {Wade}, G.~A. and 
	{Weiss}, W. and {Buysschaert}, B. and {Ramiaramanantsoa}, T. and 
	{Bratcher}, A.~D. and {Gerhartz}, C.~J. and {Greco}, J.~J. and 
	{Hardegree-Ullman}, K. and {Lembryk}, L. and {Oswald}, W.~L.
	},
    title = "{The most massive heartbeat: an in-depth analysis of {$\iota$} Orionis}",
  journal = {\mnras},
archivePrefix = "arXiv",
   eprint = {1703.02086},
 primaryClass = "astro-ph.SR",
     year = 2017,
    month = may,
   volume = 467,
    pages = {2494-2503},
      doi = {10.1093/mnras/stx207},
   adsurl = {http://esoads.eso.org/abs/2017MNRAS.467.2494P}
}

@ARTICLE{2021ApJ...909..149G,
       author = {{Ghoreyshi}, Mohammad R. and {Carciofi}, Alex C. and {Jones}, Carol E. and {Faes}, Daniel M. and {Baade}, Dietrich and {Rivinius}, Thomas},
        title = "{A Multi-Observing Technique Study of the Dynamical Evolution of the Viscous Disk around the Be Star {\ensuremath{\omega}} CMa}",
      journal = {\apj},
         year = 2021,
        month = mar,
       volume = {909},
       number = {2},
          eid = {149},
        pages = {149},
          doi = {10.3847/1538-4357/abdd1e},
archivePrefix = {arXiv},
       eprint = {2102.04430},
 primaryClass = {astro-ph.SR},
       adsurl = {https://ui.adsabs.harvard.edu/abs/2021ApJ...909..149G}
}

@ARTICLE{1931ApJ....73...94S,
   author = {{Struve}, O.},
    title = "{On the Origin of Bright Lines in Spectra of Stars of Class B}",
  journal = {\apj},
     year = 1931,
    month = mar,
   volume = 73,
    pages = {94},
      doi = {10.1086/143298},
   adsurl = {http://esoads.eso.org/abs/1931ApJ....73...94S}
}

@ARTICLE{2017A&A...601A..74K,
       author = {{Klement}, R. and {Carciofi}, A.~C. and {Rivinius}, T. and {Matthews}, L.~D. and {Vieira}, R.~G. and {Ignace}, R. and {Bjorkman}, J.~E. and {Mota}, B.~C. and {Faes}, D.~M. and {Bratcher}, A.~D. and {Cur{\'e}}, M. and {{\v{S}}tefl}, S.},
        title = "{Revealing the structure of the outer disks of Be stars}",
      journal = {\aap},
         year = 2017,
        month = may,
       volume = {601},
          eid = {A74},
        pages = {A74},
          doi = {10.1051/0004-6361/201629932},
archivePrefix = {arXiv},
       eprint = {1703.07321},
 primaryClass = {astro-ph.SR},
       adsurl = {https://ui.adsabs.harvard.edu/abs/2017A&A...601A..74K}
}

@ARTICLE{1997ApJ...479..477Q,
       author = {{Quirrenbach}, A. and {Bjorkman}, K.~S. and {Bjorkman}, J.~E. and {Hummel}, C.~A. and {Buscher}, D.~F. and {Armstrong}, J.~T. and {Mozurkewich}, D. and {Elias}, N.~M., II and {Babler}, B.~L.},
        title = "{Constraints on the Geometry of Circumstellar Envelopes: Optical Interferometric and Spectropolarimetric Observations of Seven Be Stars}",
      journal = {\apj},
         year = 1997,
        month = apr,
       volume = {479},
       number = {1},
        pages = {477-496},
          doi = {10.1086/303854},
       adsurl = {https://ui.adsabs.harvard.edu/abs/1997ApJ...479..477Q}
}

@ARTICLE{1975ApJ...196..777P,
       author = {{Poeckert}, R.},
        title = "{Linear polarization of Halpha in Be stars.}",
      journal = {\apj},
         year = 1975,
        month = mar,
       volume = {196},
        pages = {777-787},
          doi = {10.1086/153468},
       adsurl = {https://ui.adsabs.harvard.edu/abs/1975ApJ...196..777P}
}

@ARTICLE{1998A&A...330..243H,
       author = {{Hummel}, W.},
        title = "{On the spectacular variations of Be stars. Evidence for a temporarily tilted circumstellar disk}",
      journal = {\aap},
         year = 1998,
        month = feb,
       volume = {330},
        pages = {243-252},
       adsurl = {https://ui.adsabs.harvard.edu/abs/1998A&A...330..243H}
}

@ARTICLE{2012A&A...546A..92G,
       author = {{Gamen}, R. and {Arias}, J.~I. and {Barb{\'a}}, R.~H. and {Morrell}, N.~I. and {Walborn}, N.~R. and {Sota}, A. and {Ma{\'\i}z Apell{\'a}niz}, J. and {Alfaro}, E.~J.},
        title = "{The discovery of a shell-like event in the O-type star HD 120678}",
      journal = {\aap},
         year = 2012,
        month = oct,
       volume = {546},
          eid = {A92},
        pages = {A92},
          doi = {10.1051/0004-6361/201118725},
       adsurl = {https://ui.adsabs.harvard.edu/abs/2012A&A...546A..92G}
}

@ARTICLE{2009A&A...505..743V,
       author = {{Vink}, Jorick S. and {Davies}, B. and {Harries}, T.~J. and {Oudmaijer}, R.~D. and {Walborn}, N.~R.},
        title = "{On the presence and absence of disks around O-type stars}",
      journal = {\aap},
         year = 2009,
        month = oct,
       volume = {505},
       number = {2},
        pages = {743-753},
          doi = {10.1051/0004-6361/200912610},
archivePrefix = {arXiv},
       eprint = {0909.0888},
 primaryClass = {astro-ph.SR},
       adsurl = {https://ui.adsabs.harvard.edu/abs/2009A&A...505..743V}
}

@ARTICLE{2022ApJ...928..145M,
       author = {{Marr}, K.~C. and {Jones}, C.~E. and {Tycner}, C. and {Carciofi}, A.~C. and {Silva}, A.~C. Fonseca},
        title = "{The Role of Disk Tearing and Precession in the Observed Variability of Pleione}",
      journal = {\apj},
         year = 2022,
        month = apr,
       volume = {928},
       number = {2},
          eid = {145},
        pages = {145},
          doi = {10.3847/1538-4357/ac551b},
archivePrefix = {arXiv},
       eprint = {2202.05900},
 primaryClass = {astro-ph.SR},
       adsurl = {https://ui.adsabs.harvard.edu/abs/2022ApJ...928..145M}
}

@ARTICLE{2023ApJ...947...27O,
       author = {{Oey}, M.~S. and {Castro}, N. and {Renzo}, M. and {Vargas-Salazar}, I. and {Suffak}, M.~W. and {Ratajczak}, M. and {Monnier}, J.~D. and {Szymanski}, M.~K. and {Phillips}, G.~D. and {Calvet}, N. and {Chiti}, A. and {Micheva}, G. and {Rasmussen}, K.~C. and {Townsend}, R.~H.~D.},
        title = "{Strong Variability in AzV 493, an Extreme Oe-type Star in the SMC}",
      journal = {\apj},
         year = 2023,
        month = apr,
       volume = {947},
       number = {1},
          eid = {27},
        pages = {27},
          doi = {10.3847/1538-4357/acb690},
archivePrefix = {arXiv},
       eprint = {2301.11433},
 primaryClass = {astro-ph.SR},
       adsurl = {https://ui.adsabs.harvard.edu/abs/2023ApJ...947...27O}
}

@ARTICLE{2022ApJS..259...16W,
       author = {{Wrona}, Marcin and {Ratajczak}, Milena and {Ko{\l}aczek-Szyma{\'n}ski}, Piotr A. and {Koz{\l}owski}, Szymon and {Soszy{\'n}ski}, Igor and {Iwanek}, Patryk and {Udalski}, Andrzej and {Szyma{\'n}ski}, Micha{\l} K. and {Pietrukowicz}, Pawe{\l} and {Skowron}, Dorota M. and {Skowron}, Jan and {Mr{\'o}z}, Przemek and {Poleski}, Rados{\l}aw and {Gromadzki}, Mariusz and {Ulaczyk}, Krzysztof and {Rybicki}, Krzysztof},
        title = "{The OGLE Collection of Variable Stars: One Thousand Heartbeat Stars in the Galactic Bulge and Magellanic Clouds}",
      journal = {\apjs},
         year = 2022,
        month = mar,
       volume = {259},
       number = {1},
          eid = {16},
        pages = {16},
          doi = {10.3847/1538-4365/ac4018},
archivePrefix = {arXiv},
       eprint = {2109.14616},
 primaryClass = {astro-ph.SR},
       adsurl = {https://ui.adsabs.harvard.edu/abs/2022ApJS..259...16W}
}

@ARTICLE{2021MNRAS.506.4083J,
       author = {{Jayasinghe}, T. and {Kochanek}, C.~S. and {Strader}, J. and {Stanek}, K.~Z. and {Vallely}, P.~J. and {Thompson}, Todd A. and {Hinkle}, J.~T. and {Shappee}, B.~J. and {Dupree}, A.~K. and {Auchettl}, K. and {Chomiuk}, L. and {Aydi}, E. and {Dage}, K. and {Hughes}, A. and {Shishkovsky}, L. and {Sokolovsky}, K.~V. and {Swihart}, S. and {Voggel}, K.~T. and {Thompson}, I.~B.},
        title = "{The loudest stellar heartbeat: characterizing the most extreme amplitude heartbeat star system}",
      journal = {\mnras},
         year = 2021,
        month = sep,
       volume = {506},
       number = {3},
        pages = {4083-4100},
          doi = {10.1093/mnras/stab1920},
archivePrefix = {arXiv},
       eprint = {2104.13925},
 primaryClass = {astro-ph.SR},
       adsurl = {https://ui.adsabs.harvard.edu/abs/2021MNRAS.506.4083J}
}

@ARTICLE{2012ApJ...753...86T,
       author = {{Thompson}, Susan E. and {Everett}, Mark and {Mullally}, Fergal and {Barclay}, Thomas and {Howell}, Steve B. and {Still}, Martin and {Rowe}, Jason and {Christiansen}, Jessie L. and {Kurtz}, Donald W. and {Hambleton}, Kelly and {Twicken}, Joseph D. and {Ibrahim}, Khadeejah A. and {Clarke}, Bruce D.},
        title = "{A Class of Eccentric Binaries with Dynamic Tidal Distortions Discovered with Kepler}",
      journal = {\apj},
         year = 2012,
        month = jul,
       volume = {753},
       number = {1},
          eid = {86},
        pages = {86},
          doi = {10.1088/0004-637X/753/1/86},
archivePrefix = {arXiv},
       eprint = {1203.6115},
 primaryClass = {astro-ph.SR},
       adsurl = {https://ui.adsabs.harvard.edu/abs/2012ApJ...753...86T}
}

@ARTICLE{2013ApJ...766..119M,
       author = {{Miroshnichenko}, A.~S. and {Pasechnik}, A.~V. and {Manset}, N. and {Carciofi}, A.~C. and {Rivinius}, Th. and {{\v{S}}tefl}, S. and {Gvaramadze}, V.~V. and {Ribeiro}, J. and {Fernando}, A. and {Garrel}, T. and {Knapen}, J.~H. and {Buil}, C. and {Heathcote}, B. and {Pollmann}, E. and {Mauclaire}, B. and {Thizy}, O. and {Martin}, J. and {Zharikov}, S.~V. and {Okazaki}, A.~T. and {Gandet}, T.~L. and {Eversberg}, T. and {Reinecke}, N.},
        title = "{The 2011 Periastron Passage of the Be Binary {\ensuremath{\delta}} Scorpii}",
      journal = {\apj},
         year = 2013,
        month = apr,
       volume = {766},
       number = {2},
          eid = {119},
        pages = {119},
          doi = {10.1088/0004-637X/766/2/119},
archivePrefix = {arXiv},
       eprint = {1302.4021},
 primaryClass = {astro-ph.SR},
       adsurl = {https://ui.adsabs.harvard.edu/abs/2013ApJ...766..119M}
}

@ARTICLE{2004AcA....54..153P,
       author = {{Pojmanski}, G. and {Maciejewski}, G.},
        title = "{The All Sky Automated Survey. Catalog of Variable Stars. III. 12h-18h Quarter of the Southern Hemisphere}",
      journal = {\actaa},
         year = 2004,
        month = jun,
       volume = {54},
        pages = {153-179},
          doi = {10.48550/arXiv.astro-ph/0406256},
archivePrefix = {arXiv},
       eprint = {astro-ph/0406256},
 primaryClass = {astro-ph},
       adsurl = {https://ui.adsabs.harvard.edu/abs/2004AcA....54..153P}
}

@ARTICLE{2022A&A...659A..47K,
       author = {{Ko{\l}aczek-Szyma{\'n}ski}, P.~A. and {Pigulski}, A. and {Wrona}, M. and {Ratajczak}, M. and {Udalski}, A.},
        title = "{Tidally excited oscillations in MACHO 80.7443.1718: Changing amplitudes and frequencies, high-frequency tidally excited mode, and a decrease in the orbital period}",
      journal = {\aap},
         year = 2022,
        month = mar,
       volume = {659},
          eid = {A47},
        pages = {A47},
          doi = {10.1051/0004-6361/202142171},
archivePrefix = {arXiv},
       eprint = {2109.11508},
 primaryClass = {astro-ph.SR},
       adsurl = {https://ui.adsabs.harvard.edu/abs/2022A&A...659A..47K}
}

@ARTICLE{2022ApJ...928..135W,
       author = {{Wrona}, Marcin and {Ko{\l}aczek-Szyma{\'n}ski}, Piotr A. and {Ratajczak}, Milena and {Koz{\l}owski}, Szymon},
        title = "{Photometric Analysis of the OGLE Heartbeat Stars}",
      journal = {\apj},
         year = 2022,
        month = apr,
       volume = {928},
       number = {2},
          eid = {135},
        pages = {135},
          doi = {10.3847/1538-4357/ac56e6},
archivePrefix = {arXiv},
       eprint = {2109.14614},
 primaryClass = {astro-ph.SR},
       adsurl = {https://ui.adsabs.harvard.edu/abs/2022ApJ...928..135W}
}

@ARTICLE{2023A&A...678A..47B,
       author = {{Baade}, D. and {Labadie-Bartz}, J. and {Rivinius}, Th. and {Carciofi}, A.~C.},
        title = "{The historical active episodes of the disks around {\ensuremath{\gamma}} Cassiopeiae (B0.5 IVe) and 59 Cygni (B1 IVe) revisited}",
      journal = {\aap},
         year = 2023,
        month = oct,
       volume = {678},
          eid = {A47},
        pages = {A47},
          doi = {10.1051/0004-6361/202244149},
       adsurl = {https://ui.adsabs.harvard.edu/abs/2023A&A...678A..47B}
}

@ARTICLE{2023MNRAS.526..782S,
       author = {{Suffak}, M.~W. and {Jones}, C.~E. and {Carciofi}, A.~C. and {de Amorim}, T.~H.},
        title = "{Non-LTE Monte Carlo radiative transfer - III. The thermal properties of tilted and warped Be star discs}",
      journal = {\mnras},
         year = 2023,
        month = nov,
       volume = {526},
       number = {1},
        pages = {782-801},
          doi = {10.1093/mnras/stad2781},
archivePrefix = {arXiv},
       eprint = {2309.04816},
 primaryClass = {astro-ph.SR},
       adsurl = {https://ui.adsabs.harvard.edu/abs/2023MNRAS.526..782S}
}

@ARTICLE{2024ApJ...968...30R,
       author = {{Rast}, R.~G. and {Jones}, C.~E. and {Carciofi}, A.~C. and {Suffak}, M.~W. and {Fonseca Silva}, A.~C. and {Henry}, G.~W. and {Tycner}, C.},
        title = "{Evolution of the Disk in the Be Binary {\ensuremath{\delta}} Scorpii Probed during Three Periastron Passages}",
      journal = {\apj},
         year = 2024,
        month = jun,
       volume = {968},
       number = {1},
          eid = {30},
        pages = {30},
          doi = {10.3847/1538-4357/ad40a2},
archivePrefix = {arXiv},
       eprint = {2404.14504},
 primaryClass = {astro-ph.SR},
       adsurl = {https://ui.adsabs.harvard.edu/abs/2024ApJ...968...30R}
}

@INPROCEEDINGS{2026enap....2..430R,
       author = {{Rivinius}, Thomas and {Klement}, Robert},
        title = "{Classical Be stars}",
    booktitle = {Encyclop.\ of Astrophys.},
         year = 2026,
       volume = {2},
        month = jan,
        pages = {430-448},
          doi = {10.1016/B978-0-443-21439-4.00042-0},
archivePrefix = {arXiv},
       eprint = {2411.06882},
 primaryClass = {astro-ph.SR},
       adsurl = {https://ui.adsabs.harvard.edu/abs/2026enap....2..430R}
}

@ARTICLE{2026arXiv260706104B,
       author = {{Bo{\v{z}}i{\'c}}, Hrvoje and {Harmanec}, Petr and {Bro{\v{z}}}, Miroslav and {Opli{\v{s}}tilov{\'a}}, Al{\v{z}}b{\v{e}}ta and {Koubsk{\'y}}, Pavel and {Hadrava}, Petr and {Ru{\v{z}}djak}, Domagoj and {Sudar}, Davor and {Wolf}, Marek and {Zasche}, Petr and {Honsa}, Jaroslav and {{\v{Z}}{\v{d}}{\'a}rsk{\'y}}, Franti{\v{s}}ek and {Harmanec}, Adam and {Jon{\'a}k}, Juraj and {Piantschitsch}, Isabell and {Skoki{\'c}}, Ivica and {{\v{S}}vr{\v{c}}kov{\'a}}, Jana and {Vitovsk{\'y}}, Kristi{\'a}n and {Vr{\v{s}}nak}, Donik and {Zummer}, Michal},
        title = "{Refined reduction and standardisation of 53 years of UBV photometry at Hvar}",
      journal = {arXiv e-prints},
         year = 2026,
        month = jul,
          eid = {arXiv:2607.06104},
        pages = {arXiv:2607.06104},
          doi = {10.48550/arXiv.2607.06104},
archivePrefix = {arXiv},
       eprint = {2607.06104},
 primaryClass = {astro-ph.SR},
       adsurl = {https://ui.adsabs.harvard.edu/abs/2026arXiv260706104B}
}

@ARTICLE{2025A&A...698A..38B,
       author = {{Bodensteiner}, J. and {Shenar}, T. and {Sana}, H. and {Britavskiy}, N. and {Crowther}, P.~A. and {Langer}, N. and {Lennon}, D.~J. and {Mahy}, L. and {Patrick}, L.~R. and {Villase{\~n}or}, J.~I. and {Abdul-Masih}, M. and {Bowman}, D.~M. and {de Koter}, A. and {de Mink}, S.~E. and {Deshmukh}, K. and {Fabry}, M. and {Gilkis}, A. and {G{\"o}tberg}, Y. and {Holgado}, G. and {Izzard}, R.~G. and {Janssens}, S. and {Kalari}, V.~M. and {Keszthelyi}, Z. and {Kub{\'a}t}, J. and {Mandel}, I. and {Maravelias}, G. and {Oskinova}, L.~M. and {Pauli}, D. and {Ramachandran}, V. and {Rocha}, D.~F. and {Renzo}, M. and {Sander}, A.~A.~C. and {Schneider}, F.~R.~N. and {Schootemeijer}, A. and {Sen}, K. and {Stoop}, M. and {Toonen}, S. and {van Loon}, J. Th. and {Valli}, R. and {Vigna-G{\'o}mez}, A. and {Vink}, J.~S. and {Wang}, C. and {Xu}, X.-T.},
        title = "{Binarity at LOw Metallicity (BLOeM): Multiplicity properties of Oe and Be stars}",
      journal = {\aap},
         year = 2025,
        month = jun,
       volume = {698},
          eid = {A38},
        pages = {A38},
          doi = {10.1051/0004-6361/202452623},
archivePrefix = {arXiv},
       eprint = {2502.02641},
 primaryClass = {astro-ph.SR},
       adsurl = {https://ui.adsabs.harvard.edu/abs/2025A&A...698A..38B}
}

@ARTICLE{2020ApJ...903..122C,
       author = {{Cheng}, Shelley J. and {Fuller}, Jim and {Guo}, Zhao and {Lehman}, Holger and {Hambleton}, Kelly},
        title = "{Detailed Characterization of Heartbeat Stars and Their Tidally Excited Oscillations}",
      journal = {\apj},
         year = 2020,
        month = nov,
       volume = {903},
       number = {2},
          eid = {122},
        pages = {122},
          doi = {10.3847/1538-4357/abb46d},
archivePrefix = {arXiv},
       eprint = {2009.01851},
 primaryClass = {astro-ph.SR},
       adsurl = {https://ui.adsabs.harvard.edu/abs/2020ApJ...903..122C}
}

@ARTICLE{1977ApJ...213..438C,
       author = {{Conti}, P.~S. and {Ebbets}, D.},
        title = "{Spectroscopic studies of O-type stars. VII. Rotational velocities V sin i and evidence for macroturbulent motions.}",
      journal = {\apj},
         year = 1977,
        month = apr,
       volume = {213},
        pages = {438-447},
          doi = {10.1086/155173},
       adsurl = {https://ui.adsabs.harvard.edu/abs/1977ApJ...213..438C}
}

@ARTICLE{2024A&A...685A.145K,
       author = {{Koenigsberger}, G. and {Estrella-Trujillo}, D.},
        title = "{Eccentric binaries: Periastron events and tidal heating}",
      journal = {\aap},
         year = 2024,
        month = may,
       volume = {685},
          eid = {A145},
        pages = {A145},
          doi = {10.1051/0004-6361/202349075},
archivePrefix = {arXiv},
       eprint = {2404.08774},
 primaryClass = {astro-ph.GA},
       adsurl = {https://ui.adsabs.harvard.edu/abs/2024A&A...685A.145K}
}

@ARTICLE{2023NatAs...7.1218M,
       author = {{MacLeod}, Morgan and {Loeb}, Abraham},
        title = "{Breaking waves on the surface of the heartbeat star MACHO 80.7443.1718.}",
      journal = {Nature Astronomy},
         year = 2023,
        month = oct,
       volume = {7},
        pages = {1218-1227},
          doi = {10.1038/s41550-023-02036-3},
       adsurl = {https://ui.adsabs.harvard.edu/abs/2023NatAs...7.1218M}
}

@ARTICLE{2014A&A...562A..37M,
       author = {{Markova}, N. and {Puls}, J. and {Sim{\'o}n-D{\'\i}az}, S. and {Herrero}, A. and {Markov}, H. and {Langer}, N.},
        title = "{Spectroscopic and physical parameters of Galactic O-type stars. II. Observational constraints on projected rotational and extra broadening velocities as a function of fundamental parameters and stellar evolution}",
      journal = {\aap},
         year = 2014,
        month = feb,
       volume = {562},
          eid = {A37},
        pages = {A37},
          doi = {10.1051/0004-6361/201322661},
archivePrefix = {arXiv},
       eprint = {1310.8546},
 primaryClass = {astro-ph.GA},
       adsurl = {https://ui.adsabs.harvard.edu/abs/2014A&A...562A..37M}
}

@ARTICLE{2025Ap&SS.370..120R,
       author = {{Rast}, Rina G. and {Jones}, Carol E. and {Suffak}, Mark and {Labadie-Bartz}, Jonathan and {ud-Doula}, Asif and {Carciofi}, Alex C. and {Quigley}, Peter and {Neiner}, Coralie and {Drake}, Jeremy J.},
        title = "{Predicted observational effects of rapid rotation for Be stars}",
      journal = {\apss},
         year = 2025,
        month = oct,
       volume = {370},
       number = {10},
          eid = {120},
        pages = {120},
          doi = {10.1007/s10509-025-04512-w},
archivePrefix = {arXiv},
       eprint = {2510.21640},
 primaryClass = {astro-ph.SR},
       adsurl = {https://ui.adsabs.harvard.edu/abs/2025Ap&SS.370..120R}
}

@ARTICLE{2026ApJ..1006..158M,
       author = {{MacLeod}, Morgan and {Loeb}, Abraham},
        title = "{Tidal Amplitudes in the Magellanic Cloud Population of Heartbeat Stars}",
      journal = {\apj},
         year = 2026,
        month = aug,
       volume = {1006},
       number = {2},
          eid = {158},
        pages = {158},
          doi = {10.3847/1538-4357/ae80c3},
archivePrefix = {arXiv},
       eprint = {2503.17133},
 primaryClass = {astro-ph.SR},
       adsurl = {https://ui.adsabs.harvard.edu/abs/2026ApJ..1006..158M}
}

@ARTICLE{2024A&A...686A.199K,
       author = {{Ko{\l}aczek-Szyma{\'n}ski}, P.~A. and {{\L}ojko}, P. and {Pigulski}, A. and {R{\'o}{\.z}a{\'n}ski}, T. and {Mo{\'z}dzierski}, D.},
        title = "{Exploring extreme brightness variations in blue supergiant MACHO 80.7443.1718: Evidence for companion-driven enhanced mass loss}",
      journal = {\aap},
         year = 2024,
        month = jun,
       volume = {686},
          eid = {A199},
        pages = {A199},
          doi = {10.1051/0004-6361/202348104},
archivePrefix = {arXiv},
       eprint = {2309.16532},
 primaryClass = {astro-ph.SR},
       adsurl = {https://ui.adsabs.harvard.edu/abs/2024A&A...686A.199K}
}

@ARTICLE{2025A&A...699A.178Q,
       author = {{Qi}, Shitao and {Song}, Hanfeng and {Meynet}, Georges and {Maeder}, Andre and {Ekstr{\"o}m}, Sylvia and {Sciarini}, Luca and {Zhang}, Ruiyu and {Qin}, Ying and {Zhan}, Qiong},
        title = "{Reverse mass transfer and rejuvenation in the massive contact system}",
      journal = {\aap},
         year = 2025,
        month = jul,
       volume = {699},
          eid = {A178},
        pages = {A178},
          doi = {10.1051/0004-6361/202554603},
       adsurl = {https://ui.adsabs.harvard.edu/abs/2025A&A...699A.178Q}
}

@ARTICLE{2019Galax...7...83K,
       author = {{Kraus}, Michaela},
        title = "{A Census of B[e] Supergiants}",
      journal = {Galaxies},
         year = 2019,
        month = sep,
       volume = {7},
       number = {4},
          eid = {83},
        pages = {83},
          doi = {10.3390/galaxies7040083},
archivePrefix = {arXiv},
       eprint = {1909.12199},
 primaryClass = {astro-ph.SR},
       adsurl = {https://ui.adsabs.harvard.edu/abs/2019Galax...7...83K}
}

@ARTICLE{2024A&A...692A.192B,
       author = {{Blex}, Susanne and {Haas}, Martin and {Chini}, Rolf},
        title = "{The rotation rate of single- and double-lined southern O stars: Determining what increases the rotation rate in binaries}",
      journal = {\aap},
         year = 2024,
        month = dec,
       volume = {692},
          eid = {A192},
        pages = {A192},
          doi = {10.1051/0004-6361/202450176},
       adsurl = {https://ui.adsabs.harvard.edu/abs/2024A&A...692A.192B}
}

@ARTICLE{1996ApJ...469..884G,
       author = {{Gies}, Douglas R. and {Barry}, Donald J. and {Bagnuolo}, Jr., William G. and {Sowers}, Jon and {Thaller}, Michelle L.},
        title = "{Spectroscopy of the Massive Binary IOTA Orionis at Periastron}",
      journal = {\apj},
         year = 1996,
        month = oct,
       volume = {469},
        pages = {884},
          doi = {10.1086/177835},
       adsurl = {https://ui.adsabs.harvard.edu/abs/1996ApJ...469..884G}
}

@ARTICLE{1972ApL....11..119H,
       author = {{Harmanec}, P. and {Koubsk{\'y}}, P. and {Krpata}, J.},
        title = "{The Shell Star 88 Herculis: Pulsating Envelope or Binary Star?}",
      journal = {\aplett},
         year = 1972,
        month = may,
       volume = {11},
        pages = {119},
       adsurl = {https://ui.adsabs.harvard.edu/abs/1972ApL....11..119H}
}

@INPROCEEDINGS{BaadeLiege,
       author = {{Baade}, D. and {Labadie-Bartz}, J. and {Rivinius}, Th.},
        title = "{One, two, three, and a tilting frisbee: a transient state of some Be stars}",
    booktitle = {Proc. 42nd Li{\` e}ge Intern.\ Astrophys.\ Coll.\ (subm.)},
         year = 2026
         }

\begin{appendix}
\section{Acknowledgements}
\label{acknowledgments}
We thank the anonymous referee for drawing our attention to the observation by Humphreys.  This work has made use of the SIMBAD database, CDS, Strasbourg Astronomical Observatory, France, the VizieR catalogue access tool, CDS, Strasbourg Astronomical Observatory, France (DOI : 10.26093/cds/vizier), the Astrophysics Data System, funded by NASA under Cooperative Agreement 80NSSC21M0056, the BeSS database, operated at LESIA, Observatoire de Meudon, France: http://basebe.obspm.fr, and the All Sky Automated Survey (ASAS) project, developed by Dr. Grzegorz Pojma\'nski at the Warsaw University Observatory. We gratefully acknowledge the contributions of the AAVSO observer community, whose photometric data and metadata resources were used in this study and made available through the AAVSO’s scientific archives.
Funded/Co-funded by the European Union (ERC, MAGNIFY, Project 101126182). Views and opinions expressed are however those of the author(s) only and do not necessarily reflect those of the European Union or the European Research Council. Neither the European Union nor the granting authority can be held responsible for them.
\end{appendix}

\end{document}